\documentclass[12pt]{article}
\usepackage[utf8]{inputenc}
\usepackage{graphicx}
\usepackage{subcaption}
\usepackage{amsmath,amsfonts,amsthm,bm}
\usepackage[colorinlistoftodos]{todonotes}
\usepackage{multirow}
\usepackage{natbib,psfrag}
\usepackage{comment}

\newcommand{\blind}{0}

\def\mb#1{\setbox0=\hbox{$#1$}
  \kern-.025em\copy0\kern-\wd0
  \kern.05em\copy0\kern-\wd0
  \kern-.025em\raise.0em\box0}

\newcommand{\var}{\text{var}}
\newcommand{\cov}{\text{cov}}
\newcommand{\E}{\text{E}}
\newcommand{\trace}{\text{trace}}

\newcommand{\beginsupplement}{%
        \setcounter{table}{0}
        \renewcommand{\thetable}{S\arabic{table}}%
        \setcounter{figure}{0}
        \renewcommand{\thefigure}{S\arabic{figure}}%
        \setcounter{section}{0}
        \renewcommand{\thesection}{S\arabic{section}}%
     }

\begin{document}

\def\spacingset#1{\renewcommand{\baselinestretch}%
{#1}\small\normalsize} \spacingset{1}

\if0\blind
{
  \title{\bf Interpretable Principal Components Analysis for Multilevel Multivariate Functional Data, with Application to EEG Experiments}
  \author{Jun Zhang \\
    Department of Biostatistics, University of Pittsburgh\\
    Greg J. Siegle \\
    Department of Psychiatry, University of Pittsburgh\\
    Wendy D'Andrea \\
    Department of Psychology, New School for Social Research \\
    and\\
    Robert T. Krafty\thanks{
    Corresponding author Robert T. Krafty, Department of Biostatistics, University of Pittsburgh, Pittsburgh, PA 15261 (e-mail:rkrafty@pitt.edu). This work is supported by National Institutes of Health grants R01GM113243 and R01MH096334.} \\
    Department of Biostatistics, University of Pittsburgh}
  \maketitle
} \fi

\if1\blind
{
  \bigskip
  \bigskip
  \bigskip
  \begin{center}
    {\LARGE\bf  Interpretable Principal Components Analysis for Multilevel Multivariate Functional Data, with Application to EEG Experiments}
\end{center}
  \medskip
} \fi

\spacingset{1.45} % DON'T change the spacing!
\newpage
\begin{center}
\section*{Abstract}
\end{center}
Many studies collect functional data from multiple subjects that have both multilevel and multivariate structures.
An example of such data comes from popular neuroscience experiments where participants' brain activity is recorded using modalities such as EEG  and summarized as power within multiple time-varying frequency bands within multiple electrodes, or brain regions. Summarizing the joint variation across multiple frequency bands for both whole-brain variability between subjects, as well as location-variation within subjects, can help to explain neural reactions to stimuli. This article introduces a novel approach to conducting interpretable principal components analysis on  multilevel multivariate functional data that decomposes total variation into  subject-level and replicate-within-subject-level (i.e. electrode-level) variation, and provides interpretable components that can be both sparse among variates (e.g. frequency bands) and have localized support over time within each frequency band. The sparsity and localization of components is achieved by solving an innovative rank-one based convex optimization problem with block Frobenius and matrix $L_1$-norm based penalties. The method is used to analyze data from a study to better understand reactions to emotional information in individuals with histories of trauma and the symptom of dissociation, revealing new neurophysiological insights into how subject- and electrode-level brain activity are associated with these phenomena. Supplementary materials for this article are available online.

\noindent%
{\it KEY WORDS:} Functional principal component analysis; Multilevel models; Psychological trauma; Regularization; Convex optimization.

\vspace{1cm}

\section{Introduction}

Functional principal components analysis (FPCA) is arguably one of the most popular tools for analyzing functional data, or data that can be modeled as realizations of continuous processes, such as curves and surfaces. FPCA provides low-dimensional, parsimonious measures that account for the majority of variation.  These measures allow one to overcome the high-dimensionality of functional data in order to visualize and understand variation within component process, and as a stage embedded within many procedures for quantifying associations between the continuous process and other variables.

There has been growing interest in FPCA for multiple dependent functional processes, such as when multiple curves are observed for each subject. Depending on the characteristics of the data and the scientific questions of interest,  multiple functional processes are typically treated either as multivariate functional data or as repeated measures functional data. FPCA for multivariate functional data \citep{rice1991, ramsay2005, chiou2014,  happ2018}  aims to describe joint variation of the different processes, and provides parsimonious representations of the data with one score per subject per principal component. Alternatively, FPCA for repeatedly observed curves from the same process for each subject, such as curves observed at different locations or times  \citep{crainiceanu2009, di2009, greven2010, staicu2010, zipunnikov2011, shou2015, goldsmith2015, scheffler2019}, aims to characterize  variation with a multilevel hierarchical structure (e.g. between- and within-subject level principal components), and provide level-specific scores per subject per component. We refer to this class of methods as multilevel FPCA.

An increasing number of studies collect and wish to analyze functional data that can be viewed  simultaneously as multilevel and multivariate.  A popular example of such data are from electroencephalography (EEG), which measures electrical activity from multiple electrodes, or locations, across the scalp.   Clinicians and researchers are often interested in frequency-domain analyses of EEG and consider time-varying activity within frequency bands, which provide interpretable information about underlying
neurophysiological mechanisms. The resulting data take the form of multiple curves over time, defined over multiple pre-specified frequency bands, recorded from multiple locations across the scalp.  Our motivating example, which is described in further detail in Section \ref{sec:bada}, involves the analysis of data from a study to better understand biological mechanisms associated with differential reactions to emotional information across individuals with and without histories of psychological trauma and the psychological phenomenon of dissociation, which is associated with feelings of numbness and being removed from reality.   To illustrate these data,  Figure 1 displays data from two of four time-varying frequency-band measurements of brain activity, at two of fourteen measured locations across the scalp. Such data can be considered multivariate across frequency-band measures, where we desire a summary of the joint information across the multiple frequency bands, as well as multilevel across locations, where we consider both whole-brain variability between participants as well as location-variation within participants.

Although methods for multivariate FPCA and for multilevel FPCA have been individually extensively studied, there is a dearth of methods that are able to jointly deal with both multivariate and repeatedly measured functional processes. A major contribution of this article is the introduction of a method for FPCA of multilevel multivariate functional data. To model repeatedly measured functional data, \cite{di2009} and \cite{shou2015} proposed to decompose the source of variation into different levels in an additive manner, analogous to mixed effect models with random effects replaced by random processes. We extend such multilevel decompositions to multivariate processes through latent random multivariate subject-specific processes and replicate-within-subject processes. The model assumes a separable replicate-temporal covariance structure that can account for correlation among data from the same subject, such as spatial correlation among data from different locations across the scalp. Covariance operators of the subject-specific and replicate-within-subject latent processes are estimated through method of moments (MoM). The eigenstructures of the subject-specific and replicate-within-subject covariance operators provide parsimonious measures for summarizing variability at the two levels and are the basis for the multilevel multivariate FPCA.

 \begin{figure}[t]
    \centering
    \includegraphics[scale=0.5]{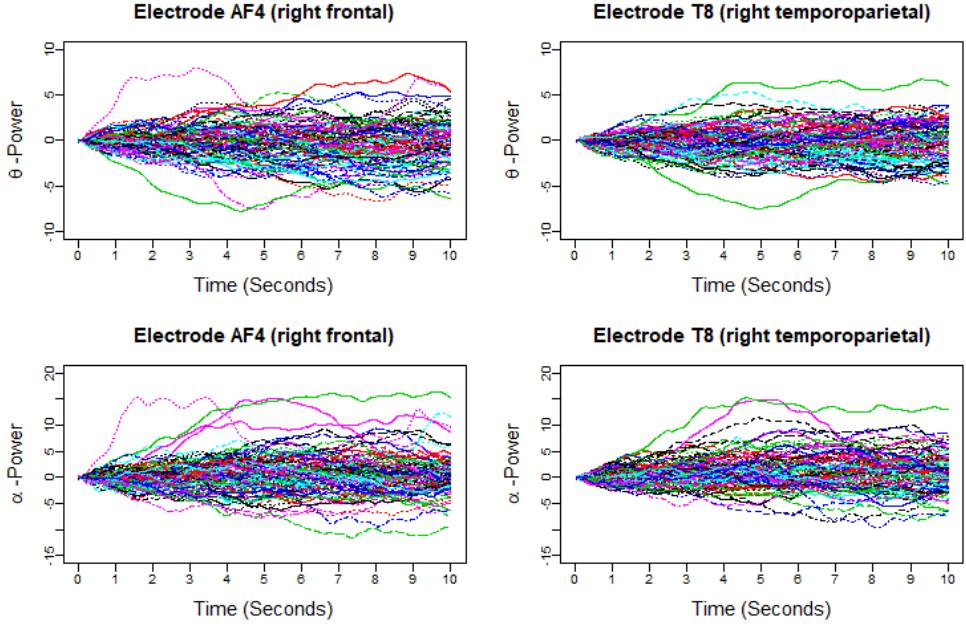}
    \caption{Illustration of BADA Study Data: time-varying power in two (theta and alpha power) of four considered bands at two (AF4, right frontal and T8, right temporoparietal) locations.}
    \label{figure 1}
\end{figure}

A second major contribution of this article is the development of an interpretable approach to the PCA of multiple functional processes that is localized both within time and among the variates. Existing approaches to conducting FPCA on multivariate functional data are limited in that they are not localized, or that each component is a nontrivial function of each variate at all time points. This lack of localization, which is a common issue across most classical procedures for analyzing functional data, can be problematic in that it obstructs interpretation. The benefit of localized procedures, both for providing scientifically interpretable measures and for improving statistical performance, have been previously discussed in the context of functional linear regression \citep{james2009, zhao2012, zhou2013} and in the context of univariate FPCA \citep{chen2015, lin2016}. However, to the best of our knowledge, localization for the PCA of multivariate functional data has yet to be addressed. The problem of conducting an interpretable FPCA for multivariate processes is considerably more challenging than for univariate, as it needs to allow for localization not only within time, but also for localization among the variates.

Towards the goal of conducting an interpretable multivariate FPCA, we introduce a novel localized sparse-variate FPCA (LVPCA). The LVPCA incorporates matrix $L_1$ and block-wise Frobenus norm-based penalties to achieve localization within time and variate, respectively. The use of this combination of penalties can be viewed as a multivariate FPCA analogue of the vector $L_1$ and block-wise vector $L_2$ norms that are used to achieve within- and between-group sparsity in high-dimensional regression by the sparse-group Lasso \citep{simon2013}. There are two main challenges when incorporating localization or sparsity into a PCA:  direct approaches via rank-one projection matrices provide solutions that, if they exist, are  NP-hard and they provide components are not orthogonal. The LVPCA overcomes the first issue through the use of the Fantope, which is the convex hull of rank-one projection matrices and has been adapted to overcome analogous problems in localized univariate FPCA \citep{chen2015} and sparse high-dimensional PCA \citep{vu2013}. Embedding the problem into the Fantope allows the LVPCA to be formulated as a convex optimization problem. The second issue is overcome through Fantope-deflation, which assures orthogonality of successive components. The formulation as a convex optimization problem allows for the development of an alternating direction method of multipliers (ADMM) algorithm \citep{boyd2011} for efficient computation and easy implementation.  For multilevel multivariate FPCA, LVPCA is applied to the MoM estimators of the subject-specific and replicate-within-subject covariance operators to provide interpretable FPCA at each level.

The rest of the article is organized as follows.  The motivating BADA Study is introduced in Section \ref{sec:bada}.   A principal components model for multilevel multivariate functional data is presented in Section \ref{sec:model} and an accompanying estimating procedure is developed in Section \ref{sec:est}.
%The estimation procedure is presented by first discussing the estimation of between- and within-subject covariances in Section \ref{sec:mom}, then discussing the LVPCA of these covariances in Section \ref{sec:lvfpca}, followed by an algorithm for implementing the procedure in Section \ref{sec:admm}, and finally discussing the selection of tuning parameters in Sections \ref{sec:rho} - \ref{sec:tuning}.
The proposed method is used to analyze simulated data in Section \ref{sec:sim} and to analyze data from the motivating BADA Study in Section \ref{sec:real}. Finally, a discussion and concluding remarks are offered in Section  \ref{sec:disc}.

\section{The BADA Study}\label{sec:bada}
\begin{comment}
Common treatments for depression, anxiety, and borderline personality disorder  work to reduce high emotional intensity and reactivity to stimuli (e.g. irritability) \citep{macleod1991}.   The opposite pole is blunted affect, or emotional blunting, in which patients fail to express emotions when confronted with emotional information.    This phenomena, often refereed to ``shutting down,'' ``dissociating,'' or ``numbing,''  is frequently observed by clinicians. However, in contrast to high emotional intensity and reactivity, blunted affect is rarely researched and is not explicitly addressed in common treatments.   This oversight could be a contributing factor as to why only 40-60\% of patients reach remission.  The Blunted and Discordant Affect (BADA) Study was conducted
% RTK: Commented out for blinded review
%at the University of Pittsburgh
to better understand biological mechanisms behind blunted affect and to inform the development of treatments.
\end{comment}

Data were examined from the Blunted and Discordant Affect (BADA) study, which was conducted to better understand individual differences in emotional information processing. Multiple psychopathologies such as depression and post traumatic stress disorder (PTSD) are characterized by intense repetitive negative self thoughts (rumination), which is increasingly well understood. These same conditions are also characterized by blunted emotional reactions and feelings of distance from the self and reality (dissociation) in response to negative information, particularly in the presence of chronic trauma or abuse backgrounds. The neural basis of such blunted reactions are less well understood. As treatments for depression and PTSD are commonly devoted to reducing negative emotion, if some individuals are already neurally disengaged or blunted in their responses, this traditional approach may not be beneficial. Thus, the BADA study examined individuals selected for a variety of psychopathologies, including those with and without chronic trauma, on tasks that could yield with blunted reactions in vulnerable individuals.

We consider data from $N=106$ study participants whose brain activity was recorded via EEG  while ruminating on a negative thought for 10 seconds.  The EEG montage included 14 electrodes placed in selected locations on the scalp (Figure \ref{fig:brain_BADA}, panel A).
%using a wireless rigid frame system (Emotiv Inc.).
Data were recorded at 128 Hz.   Frequency-band measures, or the amount of variability within an EEG time series due to osculations within an interval of frequencies, provide interpretable measures that are used by researchers and clinicians to elucidate neurophysiological mechanisms. Frequency-band measures are not independent and rather, often have high correlation. We consider four measures: theta power between 4-7 Hz, which has been shown to be linked to memory and emotional regulation \citep{knyazev2007}, alpha power between 8-12 Hz, which has been shown to often reflect relaxation, disengagement, or a lack of cognitive activity \citep{davidson1990}, beta power between 18-25 Hz, which has been linked to a variety of attentional processes \citep{neuper2009}, and gamma power between 39-45 Hz, which is associated with feature integration and fundamental cognitive processes \citep{tallon1999}.  A continuous Morlet wavelet transformation was applied and used to compute time-varying power within each of the four frequency bands; additional technical details with regards to data processing are provided in Supplemental Material. To illustrate these data, Figure 1 displays time-varying theta and alpha power from two electrodes from the 106 participants.

We desire an analysis of these data to address three questions.  First, we desire low-dimensional measures that can be used to describe variability in neurophysiological reactivity in participants while ruminating on negative thoughts.
%% Greg Insert
We are specifically interested in neurally meaningful phenomena that could happen anywhere in the brain (e.g., blunting, which occurs when brain reactivity ends before it might be expected to, or feature binding, associated with gamma-band EEG \citep{tallon1999}, which could occur in different topographies for cognitive, visual, or auditory features). This would require determination of processes that could vary anywhere throughout the brain (i.e., are not spatially localized) and across participants, but have unique time and frequency-band characteristics.
%% End insert
Second, we desire an understanding of the association between these measures with clinical measures of dissociation in order to better understand neurophysiological mechanisms behind blunted emotional reactions. To measure clinical dissociation, the total score on the Dissociative Experiences Scale (DES) was computed for each participant \citep{bernstein1986}.  The DES is a validated measure of lack of normal integration of thoughts, feelings and experiences into the stream of consciousness and memory. Scores range from 0-100, with higher scores representing greater dissociation.
%% Greg Insert
Given a concentration of lower scores in this sample, we applied a square-root transformation to DES scores to yield an approximately Gaussian distribution.
%% end insert
Lastly, blunted affect has been observed in individuals with a history of trauma and the mechanism driving blunted affect could be different among those with and without a history of trauma \citep{miniati2010}.  We are interested in understanding the potential role that trauma plays in moderating the relationship between dissociation and neurophysiology.

\section{Model}\label{sec:model}

The primary methodological question considered in this article is how to conduct interpretable principal component analyses that summarize the variability of multilevel multivariate functional data, $\mb{Y}_{ij}(t)=\{Y_{ij}^{(1)}(t),...,Y_{ij}^{(M)}(t)\}$, observed from $i=1,...,N$ subjects, with $j=1,...,J$ repeated measures taken for each subject at $M$ variates. In the motivating study, there are $N=106$ participants, $J=14$ electrodes and $M=4$ frequency band measures.
We consider the scenario where curves for each variate from each subject are observed over a common dense grid of $P$ time points as is typical for EEG data, $t_1<...<t_P\in\mathcal{T}$, and where the design is balanced with an equal number of repeated measures $J$ for each subject. Discussions with regards to the unbalanced design and to the setting where curves are observed either sparsely or over different time points are given in Section \ref{sec:disc}.

Consider a two-way functional ANOVA model
\begin{equation}
  \mb{Y}_{ij}(t)=\mb{\mu}(t)+\mb{\eta}_{j}(t)+\mb{Z}_{i}(t)+\mb{W}_{ij}(t) +\mb{\epsilon}_{ij}(t)
  \label{eq:2wayANOVA}
\end{equation}
where $\mb{\mu}(t)$ and $\mb{\eta}_{j}(t)$ are fixed effects. In our motivating example, they represent the overall mean function and electrode-specific shifts from the overall mean function. The random process $\mb{Z}_i(t)$ is the zero-mean subject-level deviation from the electrode-specific mean function, with a between-subject covariance function ${K}_z(t,s)=\E[\mb{Z}_i(t)\mb{Z}_i(s)^T]$. The random process $\mb{W}_{ij}(t)$ is the zero-mean correlated electrode-level deviation from the subject-level mean, with a within-subject covariance function ${K}_w(t,s)=\E[\mb{W}_{ij}(t) \mb{W}_{ij}(s)^T]$. The processes $\mb{W}_{ij}(t)$ and $\mb{Z}_i(t)$ are assumed to be uncorrelated.
%% Greg Insert:
%Whereas EEG analyses generally consider the overall topography of responses to a stimulus $\mb{\eta}_{j}(t)$, we are interested in capturing individual differences, which may vary topologically, in processes that could qualify that topography, such as blunted reactivity in some brain area, captured here in $\mb{W}_{ij}(s)^T$.
%% End insert

Within each subject, we assume that the covariance between $\mb{W}_{ij}(t)$ and $\mb{W}_{ik}(s)$ at electrode $j$ and $k$ is separable, and takes the form
$$\E\left[\mb{W}_{ij}(t) \mb{W}_{ik}(s)^T\right]=\rho_{jk} {K}_w(t,s) $$
where $\rho_{jk}$ is the  correlation coefficient between electrode-level deviations. Another interpretation for $\rho_{jk}$ is the correlation between observations at electrode $j$ and $k$ beyond the dependency accounted by the subject-level random process.  We assume that $\rho_{jk}$ is sparse such that $\rho_{jk}$ is non-zero for only a subset of pairs of electrodes.  It should be noted that this assumption is different than the assumption made in other multilevel principal component models where within-subject correlation is modeled as a vanishing stationary function of known structural distance \citep{li2007, staicu2010}.  In the analysis of EEG, data from different electrodes from the same subject are expected to be correlated not only due to the structural spatial location of the electrodes, but also due to functional relationships and networks.  Although the structural spatial distance between different electrodes are known, functional relationships are not. Consequently, we make the nonparametric assumption that $\rho_{jk}$ is sparse.

%Let $\Delta_{j,u}$ be a functional distance between electrode $j$ and $u$. Assume
%$$ \rho_{j,u}=\left\{
%\begin{array}{rcl}
%     0, & \text{if } \Delta_{j,u}\in \Delta^*  \\
%     1, & \text{if }  j=u, \Delta_{j,u}=0 \\
%     \in (0,1), & \text{otherwise}
%\end{array} \right.
% $$
%where $\Delta^*=\{\Delta_{j,u}|\Delta_{j,u}\rightarrow\infty\}$ is defined as the vanish range, so that electrode $j$ and $u$ with a distance of $\Delta^*$ can be considered with uncorrelated electrode level deviations. Here we assume $ 0 \le \rho_{j,u}\le 1$ because for even if some $\rho_{j,u}<0$, we can always reparameterize the model with a different $Z_i(t)$ and $W_{ij}(t)$ to have only zero and positive spatial correlations. \jz{Note sure here. Please see the other attachment.}\\

When the processes $\mb{Z}_i(t)$ and $\mb{W}_{ij}(t)$ are square-integrable, the Karhunen-Loeve expansion allows the model in Equation \ref{eq:2wayANOVA} to be expressed as
\begin{equation} \label{eq:fpca}
    \mb{Y}_{ij}(t)=\mb{\mu}(t)+\mb{\eta}_{j}(t)+
    \sum_{r=1}^{\infty}\xi_{i r}^z\mb{\phi}_{r}^z(t)+
    \sum_{r=1}^{\infty}\xi_{ij r}^w\mb{\phi}_{r}^w(t)  +\mb{\epsilon}_{ij}(t),
\end{equation}
where $\mb{\phi}_{r}^z(t)$ and $\mb{\phi}_{r}^w(t)$ are the $r^{th}$ eigenfunctions of ${K}_z(t,s)$ and ${K}_w(t,s)$, respectfully.   The principal component scores
$\xi_{i r}^z=\int_{t\in \mathcal{T}} \mb{Z}_i^T(t) \mb{\phi}_{r}^z(t)dt$  and $
\xi_{ij r}^w=\int_{t\in \mathcal{T}} \mb{W}_{ij}^T(t) \mb{\phi}_{r}^w(t)dt$
are mean-zero random variables with $\var(\xi_{i r}^z)=\theta_{r}^z$, $\var(\xi_{ij r}^w)=\theta_{r}^w$, $\cov(\xi_{ij r}^w,\xi_{iu r}^w)=\rho_{jk}\theta_{r}^w$ and are uncorrelated otherwise.

The goal of our analysis is to conduct a FPCA by obtaining interpretable estimates of the level-specific weight functions $\mb{\phi}^z_{r}$ and $\mb{\phi}^w_{r}$, which provide a low-dimensional representation of the major modes of variation, and of the  level-specific principal component scores $\xi_{i r}^z$ and $\xi^w_{i j r}$.   This FPCA is multivariate in that the weight functions are $M$-dimensional and describe the joint variation among the $M$-variates.  It is multilevel in that it provides subject-level components to describe subject-average variability and electrode-within-subject-level scores to describe within-subject variation.

\section{Estimation}\label{sec:est}
In this section, we develop a two-stage estimation procedure for conducting interpretable multilevel multivariate FPCA.   The first stage, which is discussed in Section \ref{sec:mom}, obtains MoM estimators of the between- and within-subject covariances, $K_z(t,s)$ and $K_w(t,s)$.   The second stage utilizes a novel penalized decomposition of the level-specific MoM estimators that produces interpretable components and weight functions that are smooth as functions of time, sparse among variates, and localized in time within variates.  The decomposition, which we refer to as localized sparse-variate functional principal component analysis (LVPCA), is introduced in Section \ref{sec:lvfpca}, and an optimization algorithm for computing it is offered in Section \ref{sec:admm}.   The first stage of the two-stage estimation procedure depends on an initial estimator of correlation among electrodes within-subjects, $\rho_{jk}$, and the second stage depends on three tuning parameters.   Details concerning the initial estimation of $\rho_{jk}$ are given in Section \ref{sec:rho} and the automated selection of tuning parameters is discussed in Section \ref{sec:tuning}.

%We develop multilevel LVPCA to efficiently estimate interpretable principal components for the functional model introduced in Section 3.1. Our approach begins by estimating the subject level and the electrode level sample covariance matrix as well as the spatial correlation between electrodes through the method of moment in Section 3.2.1. After estimating the covariance operators, instead of directly performing spectral decomposition as in most of the previous work, we develop a novel method in Section 3.2.2 to obtain interpretable eigenfunctions by incorporating matrix $L_1$ and block-wise Frobenius norm-based penalties to acheive localization within time and variate, respectively. Instead of maximizing over the space of variance which is NP-hard, we maximize over the convex hull of it, or the Fantope, to ensure convexity and computation. Such convex formulation then allows for the development of an alternating direction method of multiplier (ADMM) algorithm which is shown in Section 3.2.3. In Section 3.2.4 we introduced two approaches, cross-validation and fraction of variance method to select the three tuning parameters involved in smoothing, localization and variate-sparsity. Finally with the estimated eigenfunctions, the principal component scores can be estimated through the best linear unbiased predictor, by treating the model \eqref{2} as a linear mixed model. Details of the BLUP estimators are provided in Appendix A.

\subsection{Covariance Matrix Estimation}\label{sec:mom}

To aid computation, we introduce additional notation and vectorize values of the $m^{th}$ variate from electrode $j$ from subject $i$ as $\mb{Y}_{ij}^{(m)}=\{Y_{ij}^{(m)}(t_1),...,Y_{ij}^{(m)}(t_P)\}^T$, then concatenate these $M$ vectors to formulate a $PM\times 1$  vector $\mb{Y}_{ij}=\{\mb{Y}_{ij}^{(1)T},...,\mb{Y}_{ij}^{(M)T}\}^T, i=1,...,N, j=1,...,J$.  The $MP \times MP$ matrices $K_z$ and $K_w$ are defined using the same concatenation such that the $\left[\left(m - 1\right) P + p\right], \left[\left(\ell - 1\right) P + q\right]$  elements of $K_z$ and $K_w$ are the $\left(m, \ell \right)$ elements of $K_z(t_p,t_q)$ and $K_w(t_p,t_q)$, respectively.  Similarly we vectorize values of the $r^{th}$ eigenfunction within the $m^{th}$ variate $\mb{\phi}_{r}^{(m)}=\{ \phi_{r }^{(m)}(t_1),...,\phi_{r }^{(m)}(t_P) \}^T$, then concatenate these M vectors to obtain the $r^{th}$ eigenvector $\mb{\phi}_{r}=\{\mb{\phi}_{r}^{(1)T},...,\mb{\phi}_{r}^{(M)T}\}^T$.  Lastly, we let the $MP\times NJ$ matrix $Y=\{\mb{Y}_{11},...,\mb{Y}_{1J},...,\mb{Y}_{N1},...,\mb{Y}_{NJ}\}$.

Without loss of generality, we assume that $\mb{Y}_{ij}(t)$ has been demeaned by subtracting the electrode-specific mean, so that $\mb{\mu}(t)+\mb{\eta}_j(t)=0$, in order to focus on the estimation of $K_z$ and $K_w$.  We extend the symmetric sum MoM estimation approach of \cite{koch1968}, which was previously used for univariate, uncorrelated, multilevel functional data by \cite{shou2015}, to obtain unbiased estimators of sandwich form $Y G_z Y^T$ and $Y G_w Y^T$ for our multivariate, correlated, multilevel functional data.

We begin by noting that
$$
E\{\mb{Y}_{ij}(t)-\mb{Y}_{nk}(t)\}\{\mb{Y}_{ij}(s)-\mb{Y}_{nk}(s)\}^T\\
= \left\{
\begin{array}{rcl}
     2\left[K_w(t,s)(1-\rho_{jk})+\sigma^2I(t=s)   I\right] & \text{ if }  i=n,j\neq k\\
     2\left[K_w(t,s)+K_z(t,s)+\sigma^2I(t=s)  I\right] & \text{ if }  i\neq n.
\end{array}
\right.
$$
Given an initial unbiased estimator of the within-subject correlation $\hat{\rho}_{jk}$, which can be obtained though the procedure discussed in Section \ref{sec:rho}, define the matrices  $F_z=2(K_w+K_z+\sigma^2 I)$ and $F_w=2(c K_w+\sigma^2 I), \text{ where } c=\frac{J-\frac{1}{J}\sum_{j=1}^J\sum_{k=1}^J\hat{\rho}_{jk}}{J-1}$.  The explicit MoM estimators of $F_z$ and $F_w$ are given by
\begin{equation*}
\begin{split}
      \widehat{F}_z & =\frac{1}{N(N-1)J^2}\sum_{i=1}^N\sum_{i\neq n}\sum_{j=1}^J\sum_{k=1}^J(\mb{Y}_{ij}-\mb{Y}_{nk})(\mb{Y}_{ij}-\mb{Y}_{nk})^T\\
               & =\frac{1}{N(N-1)J^2}Y(J I_{NJ}-\mb{1}_{NJ}\mb{1}_{NJ}^T-B+E^TE)Y^T\\
    \widehat{F}_w & =\frac{1}{NJ(J-1)}\sum_{i=1}^N\sum_{j=1}^J\sum_{k\neq j}(\mb{Y}_{ij}-\mb{Y}_{ik})(\mb{Y}_{ij}-\mb{Y}_{ik})^T\\
               & =\frac{1}{NJ(J-1)}Y(B-E^TE)Y^T,\\
\end{split}
\end{equation*}
where $B=I_N\bigotimes JI_J$, $E=I_N\bigotimes \mb{1}_J$, $\mb{1}_J=(1,1,...,1)^T$ is the vector of ones of length $J$, and $I_J$ is the $J \times J$ identity matrix.  The covariance estimators can then be obtained as
\begin{equation*}
\widehat{K}_z  = \frac{1}{2}\widehat{F}_z - \frac{1}{2c}\widehat{F}_w, \quad
\widehat{K}_w  = \frac{1}{2c}\widehat{F}_w.
\end{equation*}
If there is no measurement error, $\widehat{K}_z$ and $\widehat{K}_w$ are consistent estimators of $\widehat{K}_z$ and $\widehat{K}_w$.  In the presence of noise, as in our case, the off-diagonal terms are consistent estimators, but there exists a nugget effect on the diagonal.  This nugget effect will be explicitly accounted for through a smoothing penalty when estimating eigenvectors.

\subsection{LVPCA: Localized Sparse-Variate Functional Principal Component Analysis}\label{sec:lvfpca}
In the second stage, we obtain interpretable principle component and weight function estimates at each level via LVPCA individually for the matrices $\widehat{K}_z$ and $\widehat{K}_w$ obtained in the first stage.  To ease notation, in this and following sections, we will use $K$ and $\mb{\phi}$ to represent an estimated covariance function and its eigenvector, which can represent either the between-subject quantities $\widehat{K}_z$ and $\widehat{\mb{\phi}}^z$ or the within-subject $\widehat{K}_w$ and $\widehat{\mb{\phi}}^w$.

\subsubsection{Methodological Motivation} \label{sec:back}
As previous discussed, many approaches for conducting a multivariate FPCA have been developed. To motivate the proposed LVPCA, here we discuss one such approach introduced by \cite{rice1991}. To produce smooth eigenvector estimates, roughness across each variate is penalized. Although any definition of roughness that can be represented as a quadratic form can be used, we consider the sum of squared second differences across time
$$ \sum_{m=1}^M\sum_{p=2}^{P-1}|{\phi}^{(m)}(t_p) - 2{\phi}^{(m)}(t_{p+1}) + {\phi}^{(m)}(t_{p+2})|^2 = \mb{\phi}^TD\mb{\phi}, $$
where $D$ is the $PM\times PM$ block diagonal matrix with $m^{th}$ block $D_0=Q^TQ$, and $Q$ is the $(P-2)\times P$ matrix where $Q_{pq}=1$ when $q\in\{p,p+2\}$, $Q_{pq}=-2$ when $q=p+1$, and is zero otherwise. Formally, given a tuning parameter $\beta$, the first eigenvector is estimated by maximizing
$\mb{\phi}^T K \mb{\phi}$ such that $\|\mb{\phi}\|_{12}=1$ and  $\mb{\phi}^T D \mb{\phi}\le \beta$,
where $\|\mb{\phi}\|_{12}=(\mb{\phi}^T \mb{\phi})^{1/2}$ is the $L_2$ vector norm and $\beta$ is a tuning parameter that controls the smoothness of the estimated component. Higher order eigenvectors are similarly defined, but with the added restriction that they are orthogonal to lower order eigenvectors. It should be noted that this approach provides estimates at the observed values, which can be interpolated to obtain eigenfunction estimates across all of $\mathcal{T}$.

It will be advantageous to consider two additional formulations of this estimator. The first, which was used by \cite{rice1991} and allows for the estimator to be computed from a simple singular value decomposition, can be obtained through Lagrange multipliers by maximizing the equivalent problem $\mb{\phi}^T(K-\gamma D)\mb{\phi}$ such that $\|\mb{\phi}\|_{12}=1$ for some smoothing parameter $\gamma>0$. Alternatively, this problem can also be expressed as maximizing
$\langle K-\gamma D, \mb{\phi}\mb{\phi}^{T}\rangle_{22}$  such that  $\|\mb{\phi}\|_{12}=1$,
where $\langle A,B\rangle_{22}=\trace(A^TB)$  is the Frobenius inner product. This last formulation will facilitate the convex relaxation of the problem when localization is introduced, assuring the existence of a solution and enabling efficient computation.

The estimated eigenfunctions previously described are not localized, either in time or among variates, in the sense that the estimated eigenfunction at every time point within each variate is non-zero with probability 1. An intuitive approach for obtaining localized estimates of the first eigenfunction is to, in a manner similar to the sparse-group Lasso \citep{simon2013}, use a combination of $L_1$ Lasso and $L_2$ group-Lasso penalties to maximize
$$  \langle K-\gamma D, \mb{\phi}\mb{\phi}^T \rangle_{22}-\alpha\sum_{m=1}^M\sqrt{P}\|\mb{\phi}^{(m)}\|_{12}-\lambda\|\mb{\phi}\|_{11} \text{  s.t.  } \|\mb{\phi}\|_{12}=1$$
where $\|\mb{\phi}\|_{11} = \sum_{m=1}^M \sum_{p=1}^P \left| \phi^{(m)} (t_p) \right|$ is the $L_1$ vector norm. The tuning parameters $\lambda>0$ and $\alpha>0$  control the degree of within- and between-variate localization, respectively. A discussion about the automated selection of $\lambda$, $\alpha$ and $\gamma$ is given in Section \ref{sec:tuning}.

\subsubsection{Penalized Deflated Fantope Estimation} \label{sec:fant}

Unfortunately, the previously considered problem is non-convex, it is not clear if or when a solution exists, and if a solution existed, it would be computationally intractable.
A computationally tractable and consistent approach can be formulated through a convex relaxation.  Approaches for conducting penalized PCA that consider problems embedded
 within the convex hull of projection matrices have been explored for sparse PCA of high-dimensional multivariate data  \citep{lei2015} and for localized PCA of univariate functional data \citep{chen2015}.  Here, we extend this approach to our setting of localized sparse-varite PCA of multivariate functional data.

Rather than attempting to maximize the objective function over the space of rank-one projection matrices of the form $\mb{\phi}\mb{\phi}^T$, we can maximize an analogous objective function over the convex hull of rank-one projection matrices, or over the Fantope
$$ \mathcal{F}=\{ H: 0\leq H \leq I, \trace(H)=1 \}. $$
Formally, we define our estimate as the first eigenvector of the matrix $\widehat{H}\in \mathcal{F}$ that maximizes
\[\langle K-\gamma D, H \rangle_{22} - \alpha \sum_{m,\ell=1}^M P \|H^{(m,\ell)}\|_{22} - \lambda \|H\|_{21},\]
where $H^{(m,\ell)}$ is the $(m,\ell)^{th} P\times P$ submatrix of $H$ and $\|H\|_{21}$ is the $L_1$ matrix norm that is the sum of the absolute values of all elements.

This approach produces estimates of the first eigenfunction, but we also desire estimates of higher-order eigenfunctions and require the collection of estimated eigenfunctions to be orthogonal. This can be easily achieved within the Fantope framework via successive Fantope-deflation. The deflated Fantope around a projection matrix $\Pi$ is defined as
$$ \mathcal{D}_\Pi=\{ H: H\in \mathcal{F}, \langle H, \Pi \rangle_{22}=0 \}.$$
Formally, define $\widehat{\Pi}_0$ as the matrix of zeros and successively estimate $\widehat{\mb{\phi}}_{r}$ for each $r=1,...,R$ as
\begin{equation}
 \begin{split}
H_{r} & =\underset{H\in\mathcal{D}_{\widehat{\Pi}_{r-1}}}{\text{argmax}}\{ \langle K-\gamma D, H \rangle_{22} - \alpha \sum_{m,\ell=1}^M P\|H^{(m,\ell)}\|_{22}- \lambda \|H\|_{21} \} \\
\widehat{\mb{\phi}}_{r} & =\text{ first eigenvector of } H_{r} \\
\widehat{\Pi}_{r} & = \widehat{\Pi}_{r-1} +\widehat{\mb{\phi}}_{r}\widehat{\mb{\phi}}_{r}^T.
\end{split}
\label{10}
\end{equation}
This provides estimates of the eigenvectors.  Scores $\xi$ can then be estimated through the best linear unbiased estimator (BLUP) and eigenvalues $\theta$ through empirical moments of the scores.  Details for the estimation of scores and eigenvalues are provided in Supplemental Material.

\subsection{Optimization using ADMM} \label{sec:admm}
The first step in our procedure \eqref{10} is a convex optimization problem. However the deflated Fantope constraint makes it difficult to directly employ block-wise subgradient strategy on the penalty terms. To solve this problem, we first rewrite the objective function in the first step of \eqref{10} as a convex global variable consensus optimization:
\begin{equation*}
     \underset{H}{\text{min}} \left\{ \mathbb{I}_{\mathcal{D}_{\widehat{\Pi}_{r-1}}}(H)-\langle K-\gamma D, H \rangle_{22} + \alpha \sum_{m,\ell=1}^M P\|H^{(m,\ell)}\|_{22} + \lambda \|H\|_{21}  \right\},
\end{equation*}
where $\mathbb{I}_{\mathcal{D}_{\widehat{\Pi}_{r-1}}}$ is 0 if $H\in\mathcal{D}_{\widehat{\Pi}_{r-1}}$ and is $\infty$ otherwise. Then the ADMM algorithm can be used so that the penalty terms can be separated from the deflated Fantope constraint. The augmented Lagrangian with auxiliary parameter $\tau>0$ is of the form
\begin{equation}
\begin{split}
\label{12}
         L_{\tau}(H,A,C)=\mathbb{I}_{\mathcal{D}_{\widehat{\Pi}_{r-1}}}(H)-\langle K-\gamma D, H \rangle_{22} + \alpha \sum_{m,\ell=1}^M P\|A^{(m,\ell)}\|_{22} \\
     + \lambda \|A\|_{21} + \frac{\tau}{2}(\|H-A+C\|_{22}^2 - \|C\|_{22}^2)
\end{split}
\end{equation}
Starting from $A_{(0)}=0, C_{(0)}=0$, for each $v=0,1,2...$, we update $H$ and $A$ alternatively by minimizing \eqref{12} with respect to $H$ and $A$. Our algorithm is described in below steps:
\begin{enumerate}
    \item Update H:
    \begin{equation*}
        \begin{split}
            H_{(v+1)} & = \underset{H\in\mathcal{D}_{\widehat{\Pi}_{r-1}}}{\text{argmin}}\left\{ -\langle K-\gamma D, H \rangle_{22} + \frac{\tau}{2}\left\|H-A_{(v)}
            +C_{(v)}\right\|_{22}^2\right\}\\
           % & = \underset{H\in\mathcal{D}_{\widehat{\Pi}_{r-1}}}{\text{argmin}}  \left\|H-(A_{(v)}-C_{(v)}+\frac{K-\gamma D}{\tau})\right\|_{22}^2\\
            & = \mathcal{P}_{\mathcal{D}_{\widehat{\Pi}_{r-1}}} \left(A_{(v)}-C_{(v)}+\frac{K-\gamma D}{\tau}\right)
        \end{split}
    \end{equation*}
    where $\mathcal{P}_{\mathcal{D}_\Pi}(B):= \underset{E\in \mathcal{D}_\Pi}{\text{argmin}}\|B-E\|^2_{22}$ is defined as a Frobenius projection operator, projecting any symmetric matrix B onto the deflated Fantope $\mathcal{D}_\Pi$, and is given in closed form in Supplemental Material.
    \item Update A:
    \begin{equation*}
     A_{(v+1)} = \underset{A}{\text{argmin}} \left\{ \alpha \sum_{m,\ell=1}^M P\|A^{(m,\ell)}\|_{22} \\
     + \lambda \|A\|_{21} + \frac{\tau}{2}\|H_{(v+1)}-A+C_{(v)}\|_{22}^2 \right\}
    \end{equation*}
    Each $(m,\ell)^{th} P\times P$ block of $A_{(v+1)}$ will be obtained by
    \begin{equation*}
        A_{(v+1)}^{(m,\ell)} = \left(1-\frac{\alpha P/\tau}{\|S_{\lambda/\tau}(H_{(v+1)}^{(m,\ell)} + C_{(v)}^{(m,\ell)})\|_{22}}\right)_+ S_{\lambda/\tau} \left(H_{(v+1)}^{(m,\ell)} + C_{(v)}^{(m,\ell)}\right)
    \end{equation*}
    where $S_b(B)$ is an element-wise soft-thresholding operator: for each element $B^{[i,j]}$  inside matrix $B,$   $S_b(B)^{[i,j]}=\text{sign}(B^{[i,j]})(|B^{[i,j]}|-b)_+$.
    \item Update dual variable C:
     \[ C_{(v+1)}=C_{(v)}+H_{(v+1)}-A_{(v+1)} \]
    \item Iterate step 1-3 until:
    \[ \text{max}\left(\left\|H_{(v+1)}-A_{(v+1)}\right\|_{22}^2, \tau^2\left\|A_{(v+1)}-A_{(v)}\right\|_{22}^2 \right) \leq \omega \] for some $\omega>0$.
\end{enumerate}

\subsection{Estimation of $\rho_{jk}$} \label{sec:rho}
The MoM estimator discussed in Section \ref{sec:mom} depends on an estimator of the within-subject-between-electrode correlation $\rho_{jk}$.  Here, we consider an estimator of $\rho_{jk}$ that takes advantage both of the separability of the within-subject electrode and temporal effects, and of the sparsity of $\rho_{jk}$.  This estimator can be viewed as a modified version of the estimator considered by \cite{staicu2010} to our setting.
Define the $M \times M$ matirix
\[f_{j k}(t, s) = 2 \left\{ K_w(t,s)\left(1-\rho_{jk}\right) + \sigma^2 I(t=s) I_M \right\}, \]
which can be consistently estimated as
\[ \hat{f}_{j k} (t,s) = \frac{1}{N} \sum_{i=1}^N \left\{\bm{Y}_{ij}(t) - \bm{Y}_{ik}(t) \right\} \left\{\bm{Y}_{ij}(s) - \bm{Y}_{ik}(s) \right\}^T. \]
Note that, due to separability of the  within-subject electrode and temporal effects, we can define the function
\[ F_{j k} = \int_{t, s \in {\cal T}} \bm{1}_{M}^Tf_{j k}(t,s)\bm{1}_{M}\text{ } dt \, ds \]
which, since $F_{j k} \propto \left(1 - \rho_{j k} \right)$, provides a measure of the disassociation of electrodes in that $F_{j k}$ is large for electrodes for which $\rho_{j k} = 0$.   A set of electrodes for which $\rho_{j k}=0$ can be estimated by thresholding the estimator
\[ \widehat{F}_{j k} = \sum_{p=1}^{P} \sum_{q \ne p} \bm{1}_M^T \hat{f}_{j k}\left(t_p, t_q\right)\bm{1}_M/P(P-1).\]
For $\delta \in (0,1)$, define the set of pairs of electrodes $ \Delta = \left\{ (j,k) \mid \widehat{F}_{j k} >\text{ upper } \delta \text{ quantile of }  \widehat{F}_{j k} \right\}.$
Our goal is to identify a subset of electrodes that are uncorrelated adjusting for subject-level deviations, and not necessarily the entire set.  Consequently, $\delta$ should be selected in a conservative manner relative to the anticipated percentage of uncorrelated electrodes.  In practice we suggest plotting all the value of $\hat{F}_{jk}$ and select somewhere below the changing point of $\hat{F}_{jk}$ to decide $\delta$ (details provided in Supplemental Material).

Given this set $\Delta$, we then define
$\tilde{f}_{\Delta}(t,s) = \sum_{\left(j,k\right) \in \Delta} \hat{f}_{j k}(t,s) / \left|\Delta\right|,$
where $\left|\Delta\right|$ is the number of pairs of electrodes in $\Delta$, which is a consistent estimator of $2 K_w(t,s)$ for $t \ne s$.  Since $\tilde{f}_{\Delta}(t,s) - \hat{f}_{j k}(t,s)$ is a consistent estimator of $2 \rho_{jk}K_W(t,s)$, we can construct a consistent estimator of $\rho_{jk}$ as
\[ \hat{\rho}_{jk} = \left\{ \sum_{p=1}^P \sum_{q \ne p} \mb{1}_M^T \left[ \tilde{f}_{\Delta}(t_p,t_q) - \hat{f}_{j k}(t_p,t_q)\right] \mb{1}_M \right\} \Big/ \left\{ \sum_{p=1}^P \sum_{q \ne p} \mb{1}_M^T\left[ \tilde{f}_{\Delta}(t_p, t_q) \right]\mb{1}_M \right\}.\]

\subsection{Tuning Parameter Selection}\label{sec:tuning}
Our optimization procedure \eqref{10} involves three tuning parameters: $\gamma$ controls smoothness as a function of time, $\alpha$ controls the among variate sparsity, and $\lambda$ controls the within time localization.  We will first select a common $\gamma$ for all eigenfunctions $\mb{\phi}_{r}, r=1,...,R$ using cross-validation \citep{rice1991}, then fix $\gamma$ and select $\alpha_{r}$ and $\lambda_{r}$ sequentially using either cross-validation or fraction of variance explained, depending the goal of the analysis.

To select $\gamma$, we employ five-fold cross-validation. The $\gamma$ is chosen among a set of candidates $\gamma_s$ such that the estimated covariance  $K^{(\nu)}$ from the validation dataset, and the estimated $H^{(-\nu)}_{r=1}(\gamma,0,0)$ from the training dataset with $\alpha$ and $\lambda$ being 0, have the largest cross-validated inner product. Formally:
\begin{equation*}
    \hat{\gamma}=\underset{\gamma \in \Upsilon_1}{\text{argmax}} \sum_{\nu=1}^5 \langle H^{(-\nu)}_{r=1}(\gamma,0,0), K^{(\nu)} \rangle
\label{13}
\end{equation*}
where $\Upsilon_1$ is a candidate set of $\gamma$, for which we used a sequence between $0$ and $P$ times the largest eigenvalue of $K$.

Similarly, $(\alpha_{r}, \lambda_{r})$ as a combination can be chosen by maximizing the cross-validated inner product of $H^{(-\nu)}_{r=1}(\gamma,\alpha_{r},\lambda_{r})$ and $K^{(\nu)}$:
\begin{equation*}
    (\hat{\alpha}_{r},\hat{\lambda}_{r})=\underset{(\alpha_{r},\lambda_{r}) \in \Upsilon_{2,r}}{\text{argmax}} \sum_{\nu=1}^5 \langle H^{(-\nu)}_{r}(\hat{\gamma},\alpha_{r},\lambda_{r}), K^{(\nu)} \rangle
\label{14}
\end{equation*}
where $\Upsilon_{2,r}$ is a candidate set of $(\alpha_{r},\lambda_{r})$. Both the candidate sequences of $\alpha_{r}$ and $\lambda_{r}$ are between 0 and the $95\%$ quantile of absolute values of off-diagonal entries in $K_{r}=(I-\widehat{\Pi}_{r-1})K(I-\widehat{\Pi}_{r-1})$. In our simulation we used coordinate descent to find the maximum cross-validated inner product.

The $(\hat{\alpha}_{r},\hat{\lambda}_{r})$ found by the cross-validation approach minimizes the bias of estimating the eigenfunction $\mb{\phi}_{r}$. When $\mb{\phi}_{r}$ is truly localized either within variates or among variates, cross-validation would be a desirable approach to reveal the true level of sparsity in $\mb{\phi}_{r}$. We adopted this method in our simulation analysis in Section \ref{sec:sim}.

Often, rather than an accurate estimate that is closet to the true $\mb{\phi}_{r}$, we are more interested in an interpretable estimate $\widehat{\mb{\phi}}_{r}$ that highlights variates and time points with dominant variation, even with some sacrifice of the fraction of variance explained (FVE). The second method of choosing $(\alpha_{r}, \lambda_{r})$ is designed to provide such interpretable $\widehat{\mb{\phi}}_{r}$. Define FVE$(\mb{\phi})=\frac{\mb{\phi}^T K \mb{\phi}}{\text{totv}(K-\gamma D)}$, where $\text{totv}(K-\gamma D)$ is the sum of all positive eigenvalues of $K-\gamma D$, which is an approximation of the total variation removing the noise $\sigma^2$. Also define relative FVE as rFVE$(\alpha_{r},\lambda_{r})=\frac{\text{FVE}(\widehat{\mb{\phi}}_{r}(\hat{\gamma},\lambda_{r},\alpha_{r}))}{\text{FVE}(\widehat{\mb{\phi}}_{r}(\hat{\gamma},0,0))}$ where $\widehat{\mb{\phi}}_{r}(\hat{\gamma},\lambda_{r},\alpha_{r})$ is the $r^{th}$ estimated eigenfunction with $\hat{\gamma},\alpha_{r}$, and $\lambda_{r}$. Then we select the largest localization under the condition that rFVE is larger than some proportion $b\in (0,1]$ that one choose to guarantee:
\begin{equation}
     (\hat{\alpha}_{r},\hat{\lambda}_{r})=\underset{(\alpha_{r},\lambda_{r})\in \Upsilon_{2,r}}{\text{argmax}} \{ \alpha_{r}+\lambda_{r}: \text{rFVE}_{r}(\alpha_{r},\lambda_{r})\geq b \}
     \label{rFVE}
\end{equation}
  If there are more than one combination providing largest $\alpha_{r}+\lambda_{r}$, we will choose the combination with largest $\alpha_{r}$ to provide a more parsimonious eigenfunction. We will illustrate this method in our real data analysis in Section \ref{sec:real}.

\section{Simulation}\label{sec:sim}
\begin{figure}
    \includegraphics[scale=0.5]{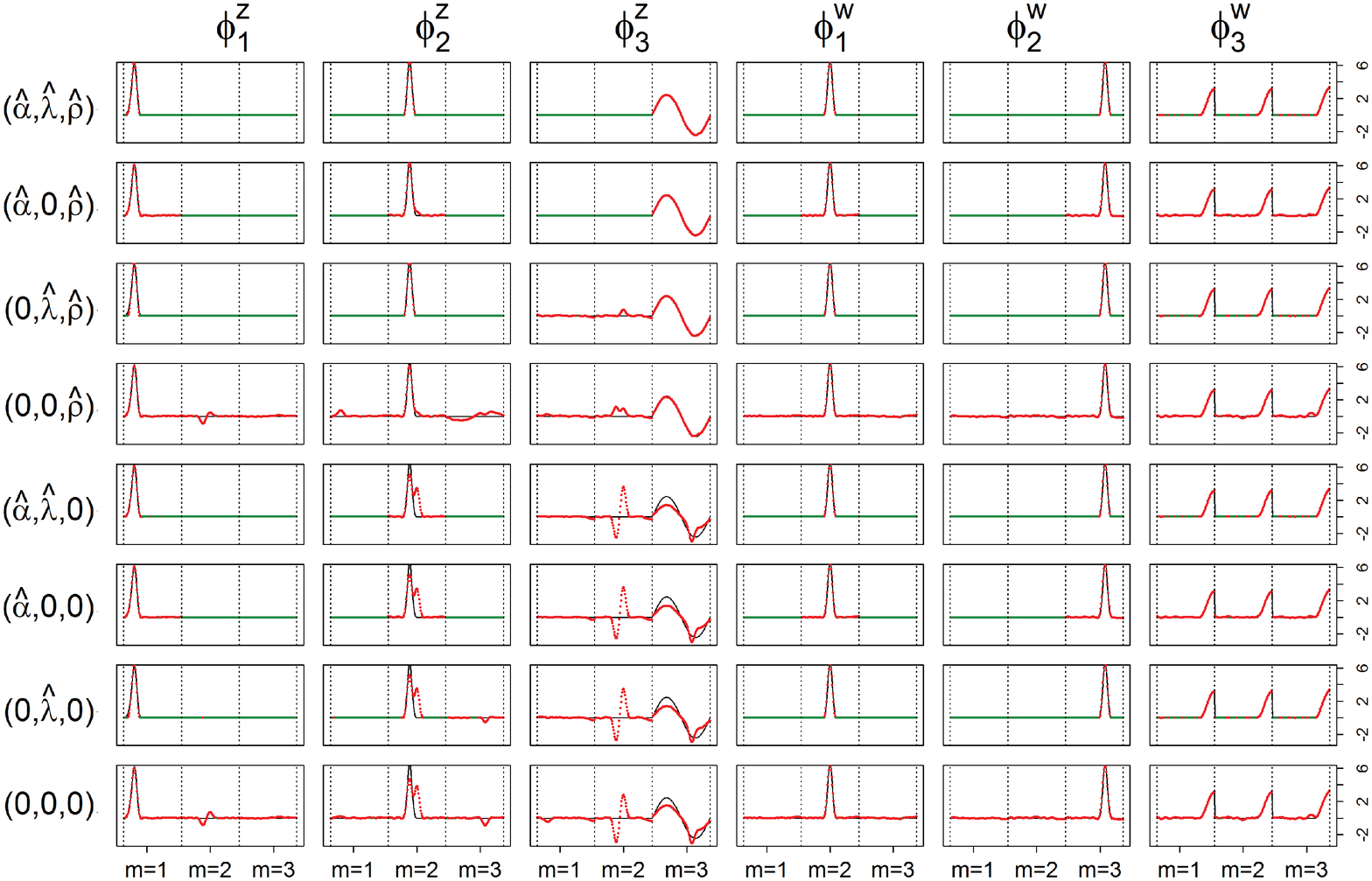}
    \caption{True (black solid) and estimated (green and red dot) eigenfunctions from one simulated data set by the described eight estimation procedures. Solid black lines are the true eigenfunctions, green lines indicate estimated zero elements, and red lines indicate estimated nonzero elements.}
    \label{figure:fit}
\end{figure}

To better understand the empirical performance of the multilevel LVPCA, we conducted simulations based on the following model:
\begin{equation}
    \mb{Y}_{ij}(t_p)= \sum_{r=1}^{R_1}\xi_{ir}^z\mb{\phi}_r^z(t_p)+
    \sum_{r=1}^{R_2}\xi_{ijr}^w\mb{\phi}_r^w(t_p)+ \mb{\epsilon}_{ij}(t_p)
\end{equation}
where $\xi_{ir}^z \sim N(0,\theta_r^z)$, $\xi_{ijr}^w \sim N(0,\theta_r^w)$, $\cov(\xi_{ijr}^w,\xi_{ikr}^w)=\rho_{jk}\theta_r^w$, $\epsilon_{ij}(t)\sim N(0, \sigma^2)$, and $t_p=(p-1)/(P-1)$, $i=1,...,N$, $j=1,...,J$, $p=1,\dots,P$.  In this section, we present results for $N=100$ subjects, $J=5$ electrodes per subject, $M=4$ variates, $P=100$ time points per variate, and $R_1=R_2=3$ eigenvalues and eigenfunctions for both subject and electrode levels.  Results for additional settings are provided in Supplemental Material. The true eigenvalues are taken as $\theta_r^z=\theta_r^w=0.5^{r-1}, r=1,2,3$, the true correlation coefficient as $\rho_{jk}=0.5$ when $j=k\pm 1$, $\rho_{jk}=0.3$ when $j=k\pm 2$, and is zero otherwise, and the noise as $\sigma^2=1$. The true eigenfuntions are specified such that $\mb{\phi}^{z}_1$, $\mb{\phi}^{z}_2$, $\mb{\phi}^{w}_1$, and $\mb{\phi}^{w}_2$ are truly sparse both within and among variates, whereas $\mb{\phi}^{z}_3$ is sparse only among variates, and $\mb{\phi}^{w}_3$ is localized only within time, as defined in Table \ref{tab:setting}.  True eigenfunctions are displayed as black lines in Figure \ref{figure:fit}.

\begin{table}
\centering
\begin{tabular}{ c c c c }
\hline
 & m=1 & m=2 & m=3 \\
\hline
$\phi^{z}_1(t)$  & $B_4(t)$ & 0 & 0\\
$\phi^{z}_2(t)$  & 0 & $B_7(t)$ & 0\\
$\phi^{z}_3(t)$  & 0 & 0 & $\sqrt{2}\sin(2\pi t)$\\
$\phi^{w}_1(t)$  & 0 & $B_9(t)$ & 0\\
$\phi^{w}_2(t)$  & 0 & 0 & $B_{12}(t)$\\
$\phi^{w}_3(t)$  & $\sqrt{2}\cos\left[\pi (t-\frac{3}{4})\right](t-\frac{3}{4})_+$ & $\sqrt{2}\cos\left[\pi (t-\frac{3}{4})\right](t-\frac{3}{4})_+$ & $\sqrt{2}\cos\left[\pi (t-\frac{3}{4})\right](t-\frac{3}{4})_+$ \\
\hline
\end{tabular}
\caption{\label{tab:setting} True eigenfunctions in the simulation setting presented in Section \ref{sec:sim}.  The function $B_b(t)$ is the $b$th cubic B-spline basis on $[0,1]$, with 16 equally spaced interior knots,  and $(t)_+ = t$ when $t \ge 0$ and is zero otherwise.}
\end{table}

Empirical performance was assessed through the square error of estimated eigenfunctions, of estimates of eigenvalues, and of the specificity (proportion zero elements correctly estimated as zero) and sensitivity (proportion of nonzero elements estimated as nonzero) of detecting nonzero values of eigenfunctions from 200 simulated random samples. To evaluate the relative contributions of the localization and sparsity penalties, and of the estimation of within-subject correlation, we consider eight estimation procedures:

\begin{enumerate}
    \item $(\alpha,\lambda, \rho)=(\hat{\alpha},\hat{\lambda}, \hat{\rho})$ corresponds to the proposed LVPCA with $\alpha$, $\lambda$ and $\gamma$ selected by 5-fold cross-validation, and $\rho$ estimated as described in Section \ref{sec:rho}.
    \item $(\alpha,\lambda, \rho)=(\hat{\alpha},0, \hat{\rho})$ corresponds to a multilevel sparse-variate FPCA without localization in time, $\alpha$ and $\gamma$ are selected by 5-fold cross-validation, $\lambda$ is restricted to be zero, and $\rho$ estimated as described in Section \ref{sec:rho}.
    \item $(\alpha,\lambda, \rho)=(0,\hat{\lambda},\hat{\rho})$ corresponds to a multilevel localized FPCA without sparsity among variates,  $\lambda$ and $\gamma$ are selected by 5-fold cross-validation, $\alpha$ is restricted to be zero, and $\rho$ estimated as described in Section \ref{sec:rho}.
    \item $(\alpha,\lambda, \rho)=(0,0, \hat{\rho})$ corresponds to a multilevel smoothed FPCA without localization in time or sparsity among variates, $\gamma$ is selected by 5-fold cross-validation, $\alpha$ and $\lambda$ are restricted to be zero, and $\rho$ estimated as described in Section \ref{sec:rho}.
    \item $(\alpha,\lambda, \rho)=(\hat{\alpha},\hat{\lambda}, 0)$: similar to method 1, but without accounting for within-subject correlation between electrode specific deviations and restricting $\hat{\rho}_{j k} = 0$.
    \item $(\alpha,\lambda, \rho)=(\hat{\alpha},0,0)$ :  similar to method 2, but without accounting for within-subject correlation between electrode specific deviations and restricting $\hat{\rho}_{j k} = 0$.
    \item $(\alpha,\lambda, \rho)=(0,\hat{\lambda},0)$:  similar to method 3, but without accounting for within-subject correlation between electrode specific deviations and restricting $\hat{\rho}_{j k} = 0$.
    \item $(\alpha,\lambda, \rho)=(0,0, 0)$:  similar to method 4, but without accounting for within-subject correlation between electrode specific deviations and restricting $\hat{\rho}_{j k} = 0$.
\end{enumerate}
For methods $1-4$, $\delta$ was set at 30\%.  Results for varying levels of $\delta$ are provided in Supplemental Material.

Figure \ref{figure:fit} displays estimated eigenfunctions $\mb{\phi}_r^z$ and $\mb{\phi}_r^w, r=1,2,3$, from one simulated data set. The eight rows correspond to the eight methods. The solid black lines are the true eigenfunctions, green lines indicate estimated zero elements and red lines indicate estimated nonzero elements. Visually, in the top row of Figure \ref{figure:fit}, it can be seen that the proposed LVPCA, which includes within-variate localization and between-variate sparseness penalties, as well as accounting for within-subject correlation between electrode-specific deviations, leads to favorable recovery of  eigenfunctions. Removing either penalty appears to potentially increase bias and select more nonzero elements of $\mb{\phi}_r^z$ than the truth. Failing to adjust for within-subject correlation between electrode-specific deviations can negatively affect the accuracy of the shape of $\hat{\mb{\phi}}_3^z$, since the between-subject variation is contaminated by the within-subject variation, specifically $\hat{\mb{\phi}}_3^z$ is contaminated by $\hat{\mb{\phi}}_1^w$.

To quantify empirical performance in estimating eigenfunctions, in Table \ref{tab:phi}  we report the median of the errors $\|\mb{\phi}-\widehat{\mb{\phi}}\|_2$ over the 200 simulations. The proposed LVPCA  outperforms the other methods when estimating subject-level eigenfunctions. When estimating electrode-level eigenfunctions, LVPCA preformed favorably, but not uniformly better compared to  $(0,\hat{\lambda}, \hat{\rho})$ and $(\hat{\alpha},\hat{\lambda}, 0)$. These result demonstrates the advantage of the proposed method in eigenvector estimation, particularly for the higher-level of the hierarchy.

\begin{table}
\centering
\begin{tabular}{ c c c c c c c}
\hline
 & $\mb{\phi}^{z}_1$ & $\mb{\phi}^{z}_2$ & $\mb{\phi}^{z}_3$ & $\mb{\phi}^{w}_1$ & $\mb{\phi}^{w}_2$ & $\mb{\phi}^{w}_3$ \\
\hline
$(\hat{\alpha},\hat{\lambda}, \hat{\rho})$ & 0.49 (0.15) &  0.91 (0.41) &  2.18 (1.15)  &  0.34 (0.04) &  0.41 (0.04)  & 0.66 (0.10) \\
$(\hat{\alpha},0,  \hat{\rho})$ & 0.60 (0.15)  &  1.89 (0.68)  &  2.54 (1.50)  &  0.66 (0.08)  &  1.11 (0.11)  &  0.91 (0.10)\\
$(0,\hat{\lambda}, \hat{\rho})$ & 1.25 (0.31)  &  0.96 (0.44) & 3.49 (1.13) & 0.41 (0.09) & 0.41 (0.05) &  0.60 (0.09)\\
$(0,0, \hat{\rho})$ & 2.67 (0.78) &  3.89 (1.00) & 4.46 (1.23) & 1.43 (0.43)&  2.14 (0.58)&   1.57 (0.44)\\
$(\hat{\alpha},\hat{\lambda}, 0)$ & 0.48 (0.14) &  1.09 (0.62) &  23.14 (1.56) &  0.33 (0.03) & 0.42 (0.05) &  0.66 (0.10) \\
$(\hat{\alpha},0, 0)$ & 0.61 (0.15) & 3.46 (1.73) &  11.71 (5.99)  & 0.66 (0.08) &  1.11 (0.11)  &  0.90 (0.09)\\
$(0,\hat{\lambda}, 0)$ & 1.30 (0.37) &  1.19 (0.72)  & 15.15 (8.45) &  0.42 (0.09) &  0.41 (0.05) &  0.60 (0.09)\\
$(0,0, 0)$ & 3.02 (0.88) & 5.52 (1.71) &  12.73 (4.39) &  1.43 (0.43) &   2.14 (0.58) &  1.57 (0.44)\\
\hline
\end{tabular}
\caption{\label{tab:phi} Median errors $\|\mb{\phi}-\widehat{\mb{\phi}}\|_2$ for $\mb{\phi}^z_r, \mb{\phi}^w_r , r=1,2,3$, (with median absolute deviations in parenthesis) over 200 simulation runs.}
\end{table}

To quantify empirical performance in identifying areas of signal,  we report the median specificity and the median sensitivity for estimating nonzero eigenvector elements in Table \ref{table:sensitivity}. Here, specificity is defined as the proportion of zero elements that are correctly estimated as zero, and sensitivity is defined as the proportion of nonzero elements correctly estimated as nonzero.
It should be noted that, in the considered setting, there is a higher proportion of zero elements compared to nonzero. For example, for $\mb{\phi}_1^z(t_p)$, there are 25 non-zero elements and 275 zero elements. The proposed LVPCA has the highest specificity among all the methods, and although not the highest, a reasonable level of sensitivity. Methods $(\hat{\alpha},0, \hat{\rho})$ and $(0,\hat{\lambda}, \hat{\rho})$ fail to localize $\mb{\phi}_3^w$ and $\mb{\phi}_3^z$ with specificities of 0.01 and 0.03, respectively. Other methods have either lower specificity or lower sensitivity than the LVPCA. This result demonstrates the advantage of the proposed LVPCA, which combines localization and between-variate sparsity penalties, in terms of variable selection.

\begin{table}
\centering
\begin{tabular}{ c c c c c c c c c c c c c c}
\hline
 & \multicolumn{6}{c}{Specificity} &  & \multicolumn{6}{c}{Sensitivity} \\
\cline{2-7} \cline{9-14}
 & $\mb{\phi}^{z}_1$ & $\mb{\phi}^{z}_2$ & $\mb{\phi}^{z}_3$ & $\mb{\phi}^{w}_1$ & $\mb{\phi}^{w}_2$ & $\mb{\phi}^{w}_3$ &
  & $\mb{\phi}^{z}_1$ & $\mb{\phi}^{z}_2$ & $\mb{\phi}^{z}_3$ & $\mb{\phi}^{w}_1$ & $\mb{\phi}^{w}_2$ & $\mb{\phi}^{w}_3$\\
\hline
$(\hat{\alpha},\hat{\lambda}, \hat{\rho})$ & 0.99 & 0.99 & 1.00     & 0.99 & 1.00 & 0.75 & & 0.87 & 0.83 & 1.00 & 0.92 &  0.92 & 0.85\\
$(\hat{\alpha},0, \hat{\rho})$ & 0.75 & 0.73 & 1.00 & 0.73 & 0.73    & 0.01 & & 1.00 & 1.00 & 1.00 & 1.00 & 1.00 & 1.00\\
$(0,\hat{\lambda}, \hat{\rho})$ & 0.97 & 1.00 & 0.03 & 1.00 & 1.00   & 0.75 & & 0.71 & 0.83 & 1.00 & 0.76 & 0.80 & 0.85\\
$(0,0, \hat{\rho})$ & 0.02 & 0.01 & 0.02 & 0.02 & 0.02 & 0.01 & & 1.00 & 1.00 & 1.00 & 1.00 & 1.00 & 1.00\\
$(\hat{\alpha},\hat{\lambda}, 0)$  & 0.99  &  0.98  & 0.86 & 1.00 & 1.00 & 0.74 & & 0.87 & 0.88 & 0.56 &  0.88 & 0.92 & 0.85\\
$(\hat{\alpha},0, 0)$  & 0.75 & 0.73 & 0.50 &  0.73 & 0.73 & 0.01 & & 1.00 & 1.00 & 1.00 & 1.00 & 1.00 & 1.00\\
$(0,\hat{\lambda}, 0)$  & 0.97 & 0.99 & 0.82 &  1.00 & 1.00 & 0.75 & & 0.71 & 0.79 & 0.91 & 0.76 & 0.80  & 0.85\\
$(0,0, 0)$ & 0.02 & 0.01 & 0.02 & 0.02 & 0.02 &  0.01 & & 1.00 & 1.00 & 1.00 & 1.00 & 1.00 & 1.00\\
\hline
\end{tabular}
\caption{\label{table:sensitivity} Median of specificity (proportion of zero elements correctly estimated as zero) and sensitivity (proportion of nonzero elements estimated as nonzero) for $\mb{\phi}^z_r, \mb{\phi}^w_r , r=1,2,3$, over 200 simulation runs.}
\end{table}

In terms of estimation of subject- and electrode-level eigenvalues, the four methods that adjust for within-subject correlation between electrode-specific deviations can recover eigenvalues with relatively little bias; median bias for these methods ranged from -0.006 to 0.022. On the contrary, the four methods without adjusting for within-subject correlation over-estimate $\theta_3^z$, with median bias ranging from 0.082 to 0.088, and highly under-estimate $\theta_1^w,\theta_2^w$ and $\theta_3^w$, with median bias ranging from -0.290 to -0.072.  This can be attributed to part of the within-subject variation is mistakenly counted as between-subject variation.  Additional details including empirical performance in estimating principal component scores, noise level and within-subject correlation, boxplots of estimated eigenvalues, as well as simulation results for additional settings are provided in Supplemental Material.

\begin{comment}
\begin{figure}
    \centering
    \includegraphics[scale=0.62]{eigenvalue_043019_n100.eps}
    \caption{Boxplot of estimated eigenvalues by the eight described estimation methods. The red solid lines indicate true eigenvalues.}
    \label{figure:eigenvalue}
\end{figure}
\end{comment}

\section{Application to the BADA Study}\label{sec:real}
We applied the proposed methodology to analyze brain reactivty while ruminating on a negative thought from the participants in the BADA study described in Section \ref{sec:bada}.   EEG data considered for each of the $i=1,\dots,N=106$ study participants are of the form $Y^{(m)}_{i j}(t_p)$ for $m=1,\dots,4$ frequency bands (theta, alpha, beta and gamma), at $j=1,\dots,J=14$ electrodes (locations displayed in Figure \ref{fig:brain_BADA}.A). Data were pre-centered around each electrode at each time point to remove fixed effects.
%In order to make the four frequency bands comparable, data were standardized such that within each frequency band, the maximum variances of power within the 10-second trial are the same.
We present the results of the analysis in two stages.  First, in Section \ref{sec:pc}, LVPCA was conducted on both the between-subject and within-subject levels to elucidate variability in neurophysiolgical activity in patients while ruminating on negative thoughts.  Then, in Section \ref{sec:reg}, regression analyses were conducted using the scores obtained from the LVPCA to quantify associations between brain activity and clinical dissociation, and for assessing  moderation of this relationship by a history of trauma.

\subsection{LVPCA}\label{sec:pc}
\begin{figure}
\begin{center}
    \includegraphics[scale=0.275]{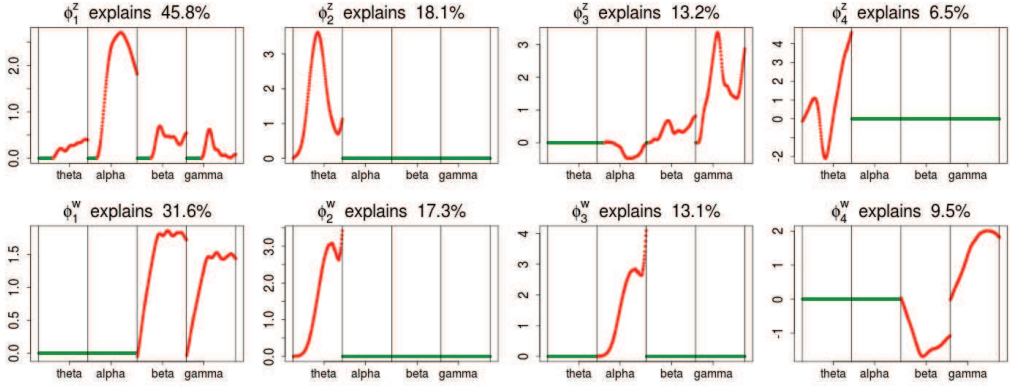}
    \caption{Estimated subject-level (top row) and electrode-level (bottom row) eigenfunctions for the BADA Study. Green lines indicate estimated zero elements and red lines indicate estimated nonzero elements.}
    \label{fig:eigenfunction_BADA}
        \end{center}
\end{figure}

LVPCA was run on MoM estimates of both within- and between-subject covariances.  The tuning parameter $\gamma$ was selected by 5-fold cross-validation, and $\alpha$ and $\lambda$ were selected based on the fraction of variance explained to maintain rFVE at $b=70\%$.   The number of principal components at each level were selected to account for $75\%$ of total variation.  The within-subject-between-electrode correlation $\rho_{jk}$ was estimated with $\delta = 20\%$.  The estimated eigenfunctions $\bm{\phi}_r^z$ and $\bm{\phi}_r^w, r=1,2,3,4$ are shown in Figure \ref{fig:eigenfunction_BADA}.

The top four subject-level principal components explain $83.6\%$ of the total between-subject variation. The first estimated subject-level eigenfunction $\mb{\phi}_1^z(t)$ explains $45.8\%$ of the variation. It is localized within each variate, being zero for each frequency band in the first 3 seconds of the trial.  It is not sparse among bands, and is a positive function of each band in the middle and end of the trial.  %Although it is not sparse among variates, the alpha band has the largest magnitude.  Consequently, the first subject-level component is a measure of total brain power in the middle and end of the trial, with an emphasis on alpha power.
The second estimated subject-level eigenfunction $\mb{\phi}_2^z(t)$, accounting for $18.1\%$ of the variation, is sparse among bands and is only a function of theta power.  The second subject-level component is a measure of total theta power, with greatest emphasis given to power in the middle of the trial.
The third estimated subject level component $\mb{\phi}_3^z(t)$ is a contrast between beta and gamma power compared to alpha power.  The fourth estimated subject-level eigenfunction $\mb{\phi}_4^z(t)$ is sparse among bands and is only a measure of theta power.   This fourth component is a contrast in theta power at the beginning and end of the trial compared to the middle.

Subject-level eigenfunctions represent the major directions of subject-specific deviation from the overall mean power function.  It is not entirely unexpected that they are largely driven by theta and alpha power. Participants were observed while ruminating on a negative memory. Theta power has been linked to memory \citep{knyazev2007}, and alpha power has been associated with a lack of emotional regulatory control \citep{klimesch2012}, two processes underlying rumination.

%% Greg insert
The power of the current method to reveal temporal variation in reactivity is illustrated as the observed time courses of reactivity are otherwise unexpected. Participants were instructed to ruminate for 10 seconds, with the expectation that brain reactivity would increase and be maintained throughout the block (as in most of the electrode-level components). Yet, most of the subject-level components show a  pattern of non-maintained reactivity, likely reflecting subject-variation in decreasing neural reactivity before the end of the 10-second block, possibly consistent with emotional blunting or disengaging from the task.
%% End insert

%% Greg insert
%\greg{Rob and I discussed putting these topologies in the supplement. See what you think of this explanation.}
The electrode-level components  represent ways individual participants vary from population mean and subject-specific whole brain responses to rumination at specific electrodes.
Estimated mean topologies of responses to the rumination instruction across all frequency bands $\widehat{\mb{\eta}}_j(t)$ are displayed in Supplemental Material.
The top four estimated electrode-level principal components explained $71.5\%$ of the total within-subject variation.  All four estimated components are sparse among frequency bands. The first estimated electrode-level eigenfunction $\mb{\phi}_1^w(t)$ is a positive function of beta and gamma power, or of high-frequency power. Rapidly increasing sustained high frequency activity could represent effortful cognition associated with trying to ruminate or to regulate emotional reactions. The estimated second and third electrode-level eigenfunctions $\mb{\phi}_2^w(t)$ and $\mb{\phi}_3^w(t)$ are positive functions of theta and alpha power, respectively. The fourth estimated electrode-level eigenfunction $\mb{\phi}_4^w(t)$ is a positive function of gamma and a negative function of beta power, which could reflect emotional engagement or feature binding.

%The majority of the total variability can be attributed to variability at different electrodes within subjects: an estimated $27.5\%$ of total variability is attributable to the between-subject effects, while the within-subject effect  accounts for the remaining $72.5\%$. The top four subject-level principal components together with the top four electrode-level principal components explain $75\%$ of the total variation.

\subsection{Association between Principal Component Scores and Dissociation} \label{sec:reg}
The multilevel LVPCA provided parsimonious, interpretable measures of the high-dimensional EEG data.  In the previous subsection, we investigated the estimated eigenfunctions and eigenvalues as a means of elucidating variation in neurophysiological activity. In this section, we use these measures to better understand associations between neurophysiological activity while ruminating on a negative thought and dissociation by regressing the square root transformed DES score that clinically measures dissociation onto principle component score, the history of trauma, and their interaction.

Since the principle component scores are correlated, we fit individual regression models for each principal component so that results can be interpreted marginally.  In total, there are 60 principal component scores of interest per subject:  the 4 subject-specific scores $\xi^z_{ir}$ together with the $14 \times 4 = 56$ subject-electrode-specific scores $\xi^w_{ijr}$, $r=1,\dots,4$, $j=1,\dots,14$.

\subsubsection{False Discovery Rate Control}\label{sec:groupBH}

To adjust for multiple testing, all reported p-values are adjusted to control the false discovery rate (FDR) at 0.05. We utilize the adaptive group Benjamin and Hochberg (GBH) procedure \citep{hu2010} for the 14 hypothesis tests grouped within anatomical regions for each electrode-level principal components, controlling the FDR for each component at $.00625=.05/8$.
Specifically, based on the correlation structure the 14 electrodes are grouped into 5 functionally distinct regions across the scalp: the right frontal (AF4, F4, F8, FC6), the right temporoparietal (T8, P8), the left frontal (AF3, F3, F7, FC5), the left temporoparietal (T7, P7), and the occipital (01, O21) region. The proportions of true null hypothesis, which are estimated through the two-stage (TST) method \citep{benjamini2006}, are assumed to be dissimilar between the five groups and the signals are more likely to appear together in these groups.

\subsubsection{Regression model and Results}

Let $Y_i$ be the square root transformed DES score,  $V_i$ be an indicator variable for a history of trauma, and $X_i$ be a principal component score for subject $i$.  We fit the linear regression model
$ \E \left(Y_i\right)= \beta_0 + \beta_1X_{i} +\beta_2V_{i} + \beta_3X_{i}V_{i} $
through least squares individually for each of 60 principal component scores. The coefficient $\beta_1$, which quantifies the association between principle component score and square root transformed DES score among participants without a history of trauma, was not statistically significant for any of the principal component scores. The coefficient $\beta_2$, which is the main effect of trauma, was statistically significant and positive for all principle component scores adjusting for multiple comparisons.  This is not unexpected, since dissociation is more common among those with a history of trauma.
There were two principle component scores with significant interactions after adjusting for multiple comparisons: one subject-level and one electrode-level.  Table \ref{tab:est} displays estimates from the two models with significant interactions and two models with borderline significant trend of interaction effects. Estimates from all 60 models are provided in Supplementary Material, as well as scatter plots of principal component scores vs. transformed DES for the models presented in Table \ref{tab:est}.

\begin{table}
\centering
\begin{tabular}{ c c c c c c c }
\hline
 & \multicolumn{2}{c}{Score} & \multicolumn{2}{c}{Trauma} & \multicolumn{2}{c}{Score $\times$ Trauma} \\
 & $\beta_1$ (SE) & p-value & $\beta_2$ (SE) & p-value & $\beta_3$ (SE) & p-value\\
\hline
$\xi^z_2$ & 0.65 (0.56) & 0.999 & 1.66 (0.25) & $<0.001$ & -2.67 (0.89) & 0.027\\
$\xi^w_{1,\text{ O1}}$  & 1.01 (0.40) & 0.098 &  1.62 (0.25) & $<0.001$ & -1.73 (0.59) & 0.036\\
$\xi^w_{3,\text{FC6}}$ & -0.43 (0.39) & 0.27 & 1.68 (0.26) & $<0.001$ & 2.10 (0.80) & 0.082\\
$\xi^w_{3,\text{F8}}$ & -0.31 (0.40) & 0.45 & 1.74 (0.26) & $<0.001$ & 2.21 (0.84) & 0.082\\
\hline
\end{tabular}
\caption{\label{tab:est} Coefficients, standard errors and adjusted p-values from univariate models with significant interaction effects and borderline significant trend of interaction effects.}
\end{table}

Among the four subject-level principal component scores, the sole model with a significant interaction was  $\xi^z_2$.  For this component, it was estimated that having a history of trauma is significantly (adjusted p-value$<0.001$) associated with more dissociation; among patients without a history of trauma the score is not associated with dissociation, however among patients with a history of trauma higher score is significantly (adjusted p-value=0.027) associated with lower level of dissociation. More specifically, recalling that the estimated second subject-level component is a measures of whole-brain theta power, if a person has a history of trauma, the higher his/her theta power, the lower their expected dissociation level.  Theta power is associated with memory and emotional control and has been shown to be elevated in participants with a history of post traumatic stress disorder \citep{bangel2017}. This result may suggest that during instructed rumination, if a person has a history of trauma, not engaging memory related circuitry, perhaps failing to reinvoke trauma recollections, even when a person wants to, may involve dissociation. In the absence of a trauma history, dissociation during rumination may be due to processes other than intensive memory recall, and thus theta power is not so strongly related to dissociation.  However, it should be noted that whole brain theta  variability may also represent noise sources \citep{simon2017, hong2013, marije2012} from outside the brain (e.g., movement), deep, possibly subcortical generators that are unlikely to be localized via scalp topography, or effects associated with global frequency power representing interactions across widespread brain systems.
Further study is needed to confirm if the moderating effect of trauma on the relationship between dissociation and this score is due to neural processes, or the result of a confounding artifact.

%Consequently, this result suggests that blunted whole-brain theta power could be a physiological mechanism connecting trauma to blunted affect.

\begin{figure}
    \centering
    %%\textbf{P-values for the interaction effect of electrode level principal component scores across 14 electrodes }\par\medskip
    \includegraphics[scale=0.7]{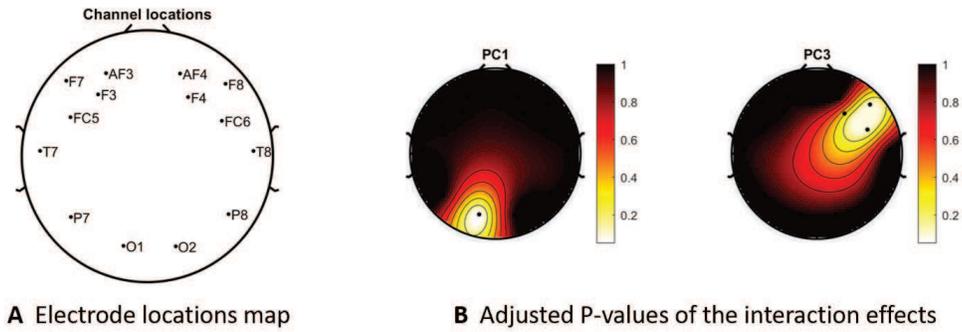}
    \caption{(A) Locations of the 14 electrodes on the brain. (B) Adjusted p-values for the interaction effect of electrode-level PC 1 and PC 3 scores across 14 electrodes displayed on their corresponding locations. White color indicates locations with p-value$<0.05$.}
    \label{fig:brain_BADA}
\end{figure}

Figure \ref{fig:brain_BADA} displays the adjusted p-values of the interaction effects for the $1$st and $3$rd electrode-level components at the 14 locations of the brain. The $2$nd and $4$th components have no effect near significant, therefore they are not shown.
The model with the $1$st principal component score $\xi^w_1$ at electrode O1 has a significant interaction effect with a history of trauma on dissociation. Specifically, among patients without a history of trauma, elevated beta and gamma power within the occipital cortex relative to the rest of the brain has a trend of being associated with increased dissociation, though the adjusted p-value is not significant (adjusted p-value=0.098). However this association is significantly different (adjusted p-value=0.036) among patients with a history of trauma, where elevated beta and gamma power within the occipital cortex relative to the rest of the brain has a negative effect on dissociation. The reason for an occipital distribution of beta and gamma power is not clear, but may suggest that failure of cognitive concentration and regulation of emotional reactions may involve dissociation among individuals with a history of trauma, while this association may have an opposite effect in the absence of trauma history.
Other scores that have different directions of associations include $\xi^w_3$ at electrode FC6 and F8. Although such differences are not significant after adjusting for multiple testing (adjusted p-value=0.082), it is worth noticing that elevated alpha power within the right frontal cortex relative to the rest of the brain has a trend of positive effect on the increased dissociation in patients with a history of trauma, which did not appear among patients without a history of trauma.

\section{Discussion} \label{sec:disc}
This article introduces a novel approach to conducting interpretable principal component analysis on multilevel multivariate functional data. The proposed localized sparse-variate FPCA (LVPCA) is combined with a multilevel covariance decomposition to provide subject-level and replicate-within-subject-level components that can be both sparse among variates as well as localized in time. The method was motivated by a study to better understand
%% Greg insert
blunted neural responses to emotional stimuli, which could occur anywhere throughout the brain,
%% end insert
to uncover connections between subject- and electrode-level brain activity that elucidates neurophysiological mechanisms connected to the phenomena of trauma patients shutting down when presented with emotional information.
%% Greg insert
Its more general application regards the ability to detect neural phenomena represented by time-frequency patterns that could occur anywhere in the brain, and thus might have different topologies for different individuals or in different tasks; blunted responding, sustained responding, and feature binding are some of the examples we have considered.
%% end insert

The proposed method can be easily reduced to useful special cases and generalized to handle more complicated data structures. By restricting $\alpha=0$ and/or $\lambda=0$, the method reduces to localized FPCA, sparse-variate FPCA or FPCA without any localization. When $J=1$ or $m=1$, the method reduces to FPCA on single level or univariate functional data. The proposed method together with all the special cases are implemented into an R package ``LVPCA'' to facilitate future research. The method can be extended by incorporating other penalty terms to take into account special structures induced by biological information \citep{beer2019}. Further, the method can be generalized from two-way to multi-way, nested or crossed study designs, as introduced in \cite{shou2015}. If the design is unbalanced with different number of repeated measures per subject, we can modify the covariance estimators by including only available $(\mb{Y}_{ij}-\mb{Y}_{ik})(\mb{Y}_{ij}-\mb{Y}_{ik})^T$ or $(\mb{Y}_{ij}-\mb{Y}_{nk})(\mb{Y}_{ij}-\mb{Y}_{nk})^T$ as each of the cross-products independently contributes a same estimator when estimating the between- and within-subject covariance.

Finally, in our motivating example, curves were observed over a common dense grid of $P$ time points. When curves are observed over different time grids, pre-smoothing can be used to obtain data over a common grid of time points \citep{ramsay2005}.  If each subject is only observed over a sparse collection of random time points, a direct smoothed covariance estimator for sparse functional data \citep{yao2005} could be used in lieu of the sample covariance  and smoothing penalization component of the proposed procedure.
%Another practical consideration regards to high-dimensional functional data. When $P$ is large, because of computation inefficiency presmoothing or prebinning on individual curves is recommended which is what we have done for the BADA study. However future work could explore more efficient ways to handle large number of time points.

\bibliographystyle{asa}
\bibliography{locstatbib}

\end{document}